\documentclass[final,5p,times,twocolumn]{elsarticle}

 \usepackage{graphicx}

\usepackage{amssymb}

\journal{Physica E}

\begin{document}

\begin{frontmatter}



\title{Selective control of edge--channel trajectories by scanning gate microscopy}


\author[label1]{N. Paradiso}
\author[label1]{S. Heun}
\author[label1]{S. Roddaro}
\author[label2]{L. N. Pfeiffer}
\author[label2]{K. W. West}
\author[label1,label3]{L. Sorba}
\author[label3]{G. Biasiol}
\author[label1]{F. Beltram}


\address[label1]{NEST, CNR--INFM and Scuola Normale Superiore, Piazza San Silvestro 12, 56127 Pisa, Italy}
\address[label2]{Bell Laboratories Lucent Technologies, 600 Mountain Avenue, Murray Hill, NJ 07974, USA}
\address[label3]{Laboratorio Nazionale TASC--INFM, Area Science Park Basovizza, 34012 Trieste, Italy}

\begin{abstract}
Electronic Mach--Zehnder interferometers in the Quantum Hall (QH) regime are currently discussed for the realization of quantum information schemes. A recently proposed device architecture employs interference between two co--propagating edge channels. Here we demonstrate the precise control of individual edge--channel trajectories in quantum point contact devices in the QH regime. The biased tip of an atomic force microscope is used as a moveable local gate to pilot individual edge channels. Our results are discussed in light of the  implementation of multi--edge interferometers.
\end{abstract}


\end{frontmatter}


\section{Introduction}
\label{Introduction}

Interference phenomena are a fundamental manifestation of the quantum mechanical nature of electrons and have promising applications in solid--state quantum information technology. Two--dimensional 
electron systems (2DES) in the quantum Hall (QH) 
regime are especially suited for this purpose given the 
large electronic coherence length characteristic of 
edge--channel chiral transport. In particular, the realization of electronic Mach--Zehnder (MZ) interferometers in QH systems appears at present a sound technology for the implementation of quantum information schemes~\cite{Ji2003}. Despite this success, the edge topology of the single--channel MZ limits the complexity of these circuits to a maximum of two interferometers~\cite{Neder2007}. In order to overcome this constraint, new device architectures were recently proposed~\cite{Giovanetti2008}, where interference paths are built using two different parallel edge channels. In this configuration, control over the interaction between the different edge channels is  challenging owing to the complexity of the edge structure.

In order to address these issues we are exploring the use of scanning gate microscopy (SGM) to control the trajectory and interaction of edge channels based on our previous results on quantum point contact (QPC) devices in the QH regime~\cite{Roddaro2003,Roddaro2004,Roddaro2005,Roddaro2009}. In the present work, the two split gates of the QPC play a double role: they not only allow us to bring the edges in close proximity, but they also provide the ability to select the edges that are sent to the center of the QPC.

\section{Experiment}
\label{Experiment}
Devices were realized starting from a high--mobility ($\mu = 2.3 \times 10^6$ cm$^2$/Vs) AlGaAs/GaAs single heterojunction with a mean free path of $\ell = 25$~$\mu$m. The two--dimensional electron system (2DES) was buried 55~nm below the surface, with a sheet density of $n = 3.2 \times 10^{11}$ cm$^{-2}$. The main results presented in this paper were reproduced with another sample of similar mobility ($\mu = 4.2 \times 10^6$ cm$^2$/Vs, $n = 1.8 \times 10^{11}$~cm$^{-2}$, 2DES  80~nm below the surface). Schottky split--gate QPCs were realized by thermal evaporation of a metal bilayer Ti/Au (10/20~nm) with a constriction gap of 300~nm. 

Experiments were performed with an atomic force microscope (AFM) operated in a $^3$He cryostat (base temperature 300~mK), equipped with a vibration and noise isolation system. The AFM has a scanning area of 8.5 $\mu$m $\times$ 8.5 $\mu$m at 300~mK. Sample temperature was 400~mK, as calibrated with a Coulomb blockade thermometer. The cryostat is equipped with a superconducting coil which provides magnetic fields of up to 9~T. 
The AFM head of our setup is constructed with a stack of actuators for both the coarse and fine control of the tip--sample position. The sample, mounted on a leadless chip carrier, is positioned on top of the piezo scanner, while the tip is connected to a tuning fork (TF). The sample topography is obtained by controlling the oscillation amplitude damping of the TF due to the tip--sample shear force (non--contact mode). Images are processed with the WSxM software~\cite{Horcas07}.

SGM conductance maps were obtained with the negatively biased AFM tip scanning above the QPC while simultaneously measuring the source--drain current through the QPC. 
The current through the device  (AC source--drain bias  100 $\mu$V) was measured using a current preamplifier in a two--probe configuration. Contact resistances were subtracted numerically. 
All measurements presented here were performed in constant--height mode (10 to 30~nm) with a tip bias of $-5$~V. This results in a capacitive coupling between tip and 2DES; no current flows from tip to sample, no strain is exerted on the sample. The biased tip acts as a local moveable gate and creates a local depletion of the 2DES under the tip.

\section{Results and discussion}
\label{Results}

\begin{figure*}[t]
	\includegraphics[width=\textwidth]{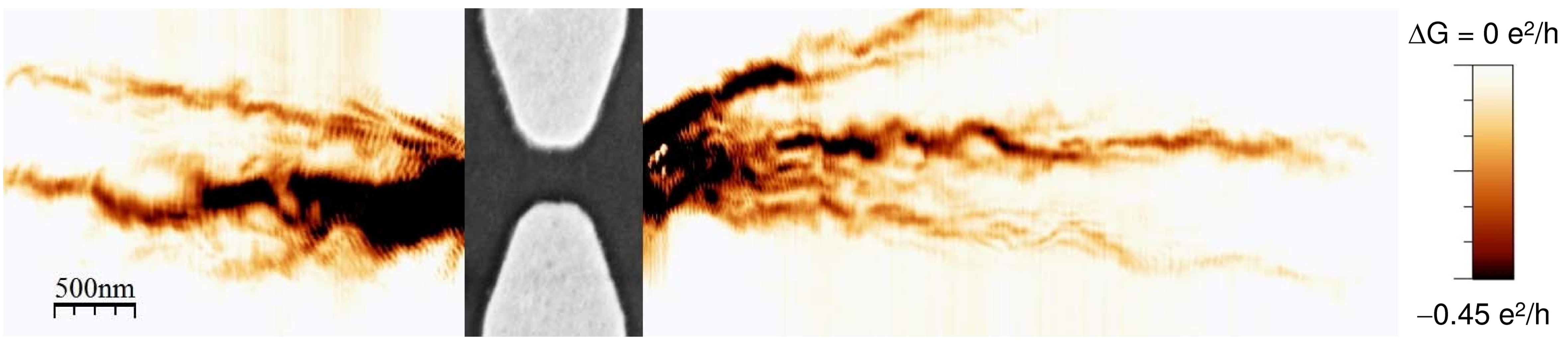}
	\caption{Characteristic branched flow observed in zero--field SGM measurements. The image shows the change in conductance $\Delta G$  as a function of tip position. The dark regions in the color plot (low conductance) correspond to the actual electron path and depend on the details of the local potential. The fringes which decorate the branches are a signature of the electron phase coherence. The center part of the image shows a scanning electron microscopy image of the QPC. Tip bias $-5$~V, tip height 10~nm, QPC conductance $G = 6e^2/h$ ($3^{rd}$ plateau), B = 0 T.}
	\label{fig:1}
\end{figure*}

\begin{figure}[t]
	\includegraphics[width=\columnwidth]{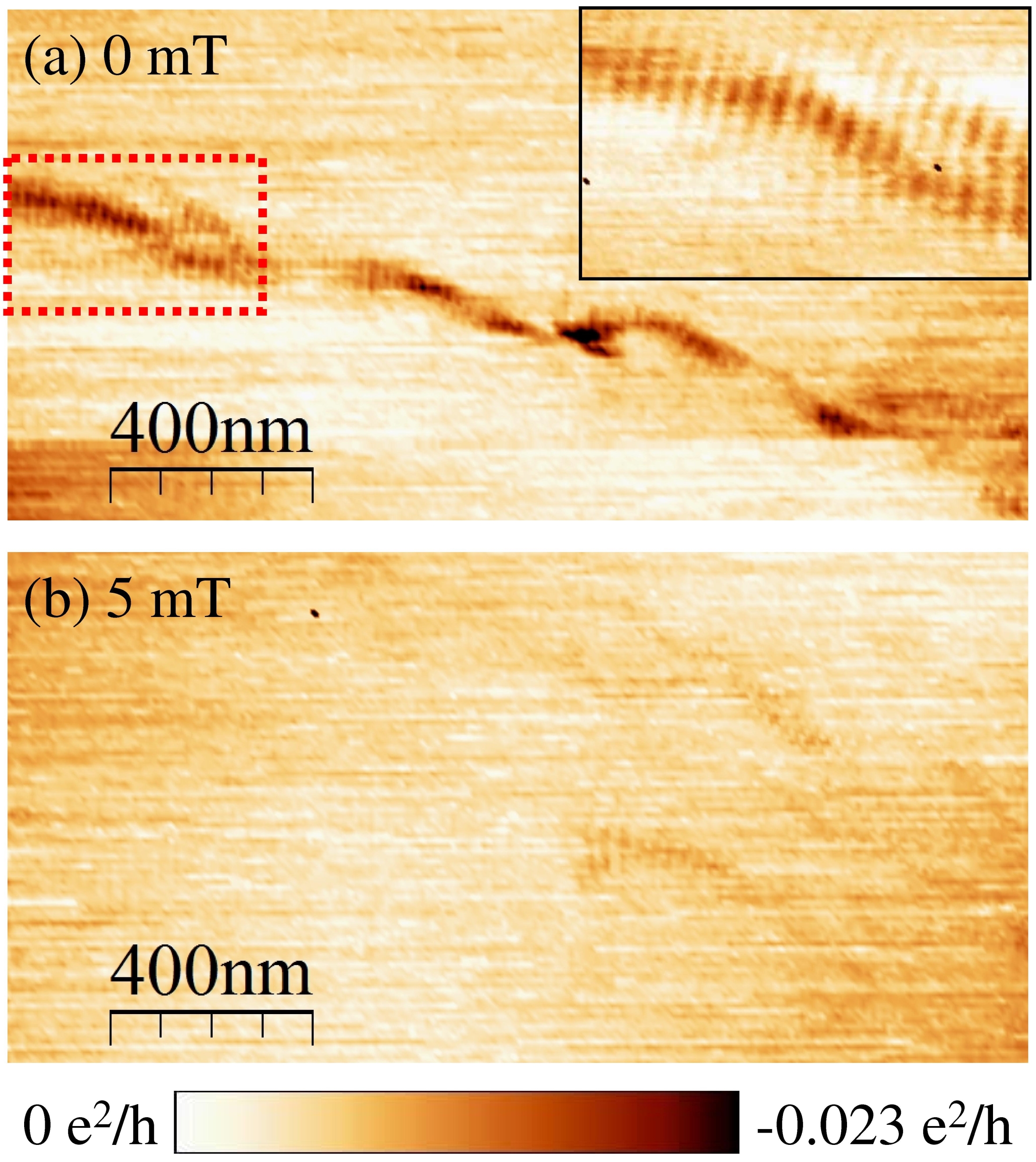}
	\caption{High--resolution images of a branch of electron flow. Images show the change in conductance $\Delta G$ as a function of tip position. The inset of (a) shows a magnified image (500~nm $\times$ 300~nm) of the area indicated by the dashed rectangle. Magnetic field in (a) 0 mT, in (b) 5 mT. Applying a magnetic field destroys the time--reversal symmetry; the branches and fringes disappear. Tip bias $-5$~V, tip height 10~nm, QPC conductance $G = 6e^2/h$ ($3^{rd}$ plateau).}
	\label{fig:2}
\end{figure}

In order to validate our microscope setup we performed SGM measurements on QPCs at zero magnetic field. When the QPC conductance is quantized, the depletion spot  induced by the tip causes a backscattering of  electrons through the QPC~\cite{Topinka2000, Topinka2001}. When the tip is located above areas with high electron flow, the conductance is reduced by this effect, while it remains constant over areas with small electron flow. Scanning the tip over the sample yields therefore images of electron flow.  The conductance map in Fig.~\ref{fig:1} shows the characteristic branched flow of electrons. These branches correspond to the paths of lowest energy for the electrons moving in the  potential landscape of the 2DES. They extend over a lengthscale of about 5~$\mu$m due to the high mobility of the 2DES. The fringes decorating these structures are separated by half the Fermi wavelength. This can be better seen in Fig.~\ref{fig:2}(a) which shows a zoom--in of one of the branches. The inset shows a high--resolution image of the area indicated by the dashed rectangle in Fig.~\ref{fig:2}(a). The fringes are clearly resolved. They are due to the interference between time--reversed electron paths from the QPC and from the depletion spot generated by the AFM tip. This can be easily verified by breaking the time reversal symmetry: when a small magnetic field (1 mT) is applied, all  fringes are washed out (see Fig.~\ref{fig:2}(b)). As already pointed out by Aidala et al.~\cite{Aidala2007}, the magnetic field also affects the electron--flow pattern:  electrons can no longer return to the QPC in a time--reversed path, because the orbits are bent, and therefore the SGM image shows constant conductance, which is independent of the tip position.

The present work is aimed at exploring the use of SGM to control the trajectory and the interaction of edge channels in the QH regime. The fundamental idea behind our experiments consists in exploiting the QPC not only to control the distance between counterpropagating edges (as in tunneling experiments~\cite{Roddaro2005, Roddaro2009}) but also to set the number of edges flowing around each of the two split gates. An essential role here is played by the SGM tip, which is exploited first to determine the filling factor underneath the two split gates and then to put the selected edge states in interaction. All the experiments discussed in the following were performed at $B=3.04$~T, which corresponds to a bulk 2DES filling factor $\nu_{bulk} = 4$. 

\begin{figure}[t]
	\includegraphics[width=\columnwidth]{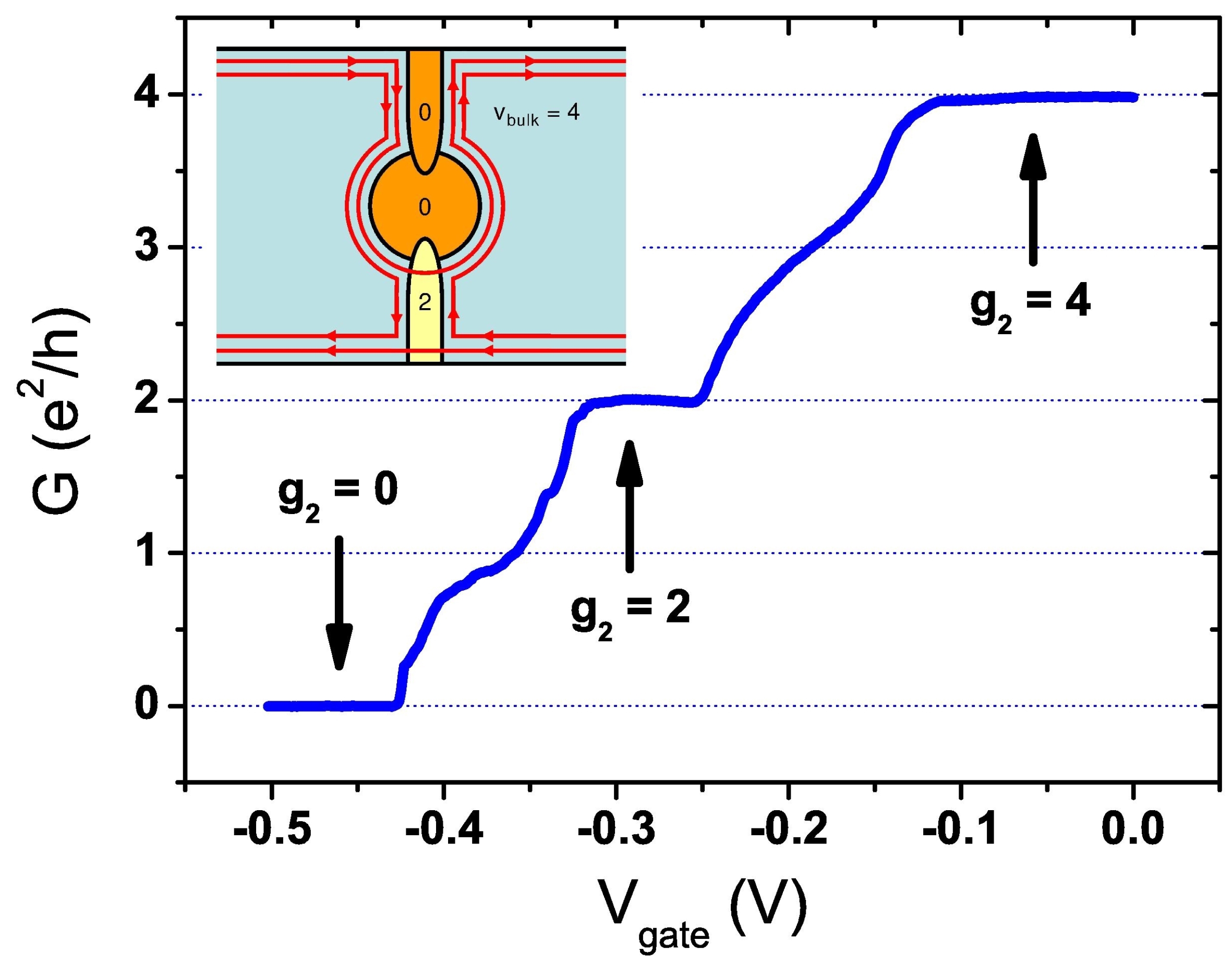}
	\caption{Source--drain conductance $G$ as a function of the voltage $V_{gate}$ applied to the lower of the two QPC gates. The inset shows a sketch of the geometry for $g_2 = 2$. A high bias voltage applied to the upper QPC gate and the tip (which is positioned above the center of the QPC) completely deplete the 2DES under these regions ($g_1 = g_{tip} = 0$). Sweeping $V_{gate}$ to negative values reduces the source--drain conductance  from the initial value to zero and allows to determine the local filling factor $g_2$ under the lower gate with high precision. B = 3.04 T, $\nu_{bulk} = 4$.}
	\label{fig:3}
\end{figure}

Figure~\ref{fig:3} shows the procedure we follow to determine the proper voltage which needs to be applied to each gate for a given filling factor. We first place the AFM tip $30$~nm above the center of the QPC. Then we apply a bias of $-5$~V to the tip to completely deplete the 2DES in the QPC gap. Next, a high bias ($-3.5$~V) is applied to one of the QPC gates, which depletes the 2DES underneath ($g_1=0$). Now, by sweeping the bias on the other gate from $0$~V to $-0.5$~V we reduce the source drain conductance from $G = \nu_{bulk} \cdot e^2/h$ down to zero. This procedure allows us to accurately determine the gate voltage corresponding to a given filling factor $g_2$ under the gate, by evaluating the position of the plateaus in the conductance graph. Because at this magnetic field the Zeeman splitting is so small, the curve in Fig.~\ref{fig:3} shows clear plateaus for even filling factors, while for odd filling factors only a shoulder is observed, i.e. we are studying spin--degenerate edge channels.

\begin{figure*}[t]
	\includegraphics[width=\textwidth]{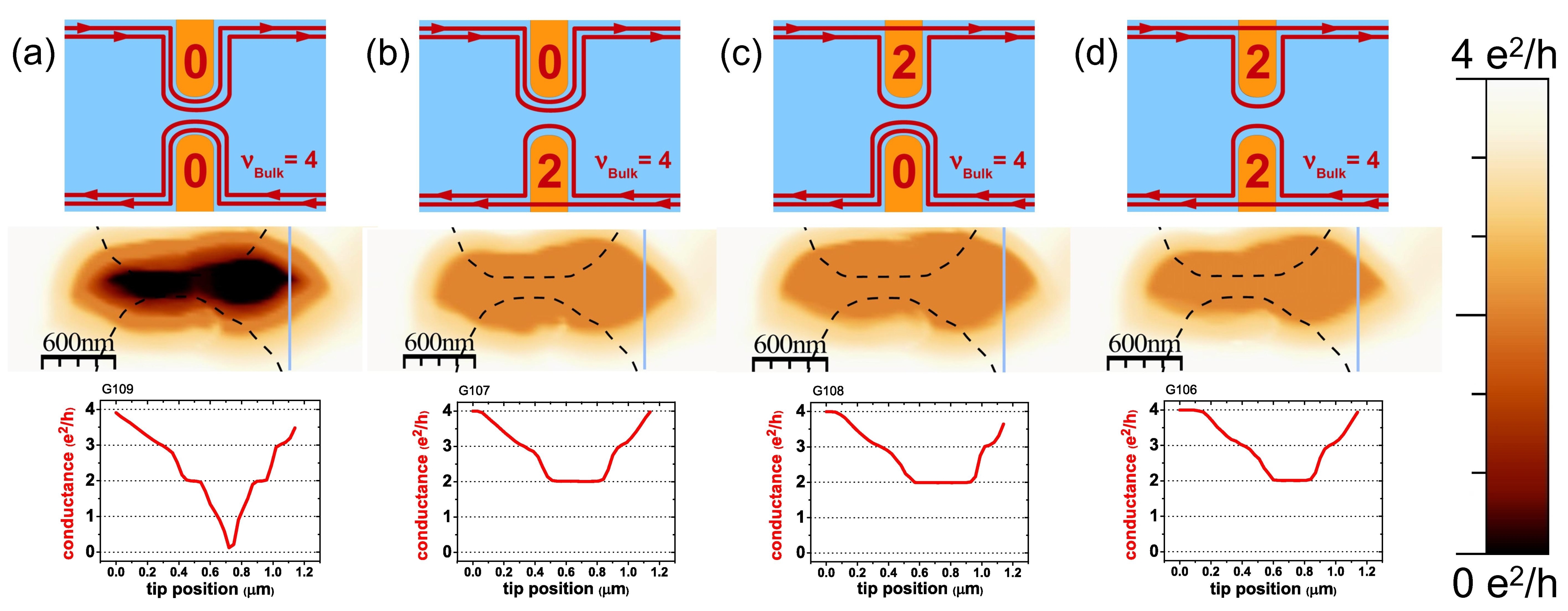}
	\caption{QPC conductance $G$ as a function of the position of the biased SGM tip. The (bulk) 2DES filling factor is set to $\nu_{bulk} = 4$ (2 spin--degenerate edge channels; $B = 3.04$~T) while the QPC gates partially or completely deplete the 2DES underneath. In (a) gate--region  filling factors $g_1 = g_2 = 0$, in (b) $g_1 = 0$, $g_2 = 2$, in (c) $g_1 = 2$, $g_2 = 0$, and in (d) $g_1 = g_2 = 2$. The top rows show sketches of the edge channel trajectories, the center row the SGM conductance images, and the lower row cross sections through the images along the vertical lines drawn in the images. The QPC outline as measured by AFM is indicated by the dashed lines. The color scale of all images ranges from $G = 4 e^2/h$ to $0 e^2/h$. Tip bias $-5$~V, tip height 30~nm.}
	\label{fig:4}
\end{figure*}

After this calibration step we performed SGM measurements in different edge--channel configurations. For instance, when the gate filling factors are set to $g_1 = g_2 = 0$, both edge channels reach the QPC center, while when $g_1 = g_2 = 2$, only one channel is deflected, whereas the other one is transmitted across the gate~\cite{Haug1988}. In total, there are four possible configurations for our SGM measurements. Figure~\ref{fig:4} shows the QPC conductance as a function of the position of the biased SGM tip ($V_{tip}=-5$~V). In the panel (a) the gate--region filling factors are $g_1=g_2=0$. When the biased tip is brought close to the QPC, pairs of edge channels are backscattered one by one, and the conductance through the QPC decreases in a step--like manner to 0. This clearly demonstrates the gating action of the tip even in the QH regime. Similar results were observed by Aoki et al. in InAlAs/InGaAs etched heterostructures with symmetric edge configuration~\cite{Aoki2005}.

Taking advantage of the high flexibility of our setup, we repeated the SGM measurements also in asymmetric configurations  such as that shown in the panel (b) of Fig.~\ref{fig:4}, where $g_1=0$ and $g_2=2$. In this case, only the pair of inner edge channels is backscattered by the local action of the tip, while the outer edge either flows far from the constriction (under the lower gate) or has no counterpart for the backscattering process to occur (upper gate). In this case, the conductance remains $G=2e^2/h$ even when the tip completely pinches off the constriction region.

When the pair of inner--edge channels is backscattered by the tip, only a \textit{single} edge channel remains in the QPC. When the tip is then moved to the center of the QPC, it completely pinches off the QPC. However, the conductance through the QPC does not change. This implies that the remaining single edge channel is moved by the tip onto a different trajectory, in this case under the lower QPC gate. This is the most important result of the present study and unequivocally demonstrates our ability to control the trajectory of individual edge channels by the tip of the SGM.

For the mirrored gate configuration in Fig.~\ref{fig:4}(c) with $g_1 = 2$ and $g_2 = 0$, we obtained  similar SGM images. We also studied the configuration depicted in Fig.~\ref{fig:4}(d) with identical filling  factors $g_1 = g_2 = 2$ under the split--gates. This situation is fundamentally different from the situations in Figs.~\ref{fig:4}(b) and (c), because here no unpaired edge channel is flowing through the QPC. The resulting SGM image, however, is similar  to those shown in  Figs.~\ref{fig:4}(b) and (c). These results clearly show that the lower bound for the conductance is determined by the number of paired edge channels which can be backscattered. 

A closer look at these images, however, reveals subtle differences between the studied cases. Looking at the upper half of the SGM images in Fig.~\ref{fig:4}, images (b) and (d) are very similar, while (c) is slightly different from the other two. On the other hand, in the lower half of the images, (c) and (d) are virtually identical, while (b) is slightly different. These observations correlate with the local filling factor under the \textit{opposite} gate: in (b) and (d), the lower gates are at $g_2 = 2$, while in (c) $g_2 = 0$; the upper gates are at $g_2 = 2$ in (c) and (d), but $g_2 = 0$ in (b). We argue that when the tip is in the upper half of the image, it moves the upper edge towards the lower edge, while it  little affects the lower edge\footnote{The lower edge is already travelling close to the lower gate and cannot be pushed further down.}. Once the upper edge approaches the lower edge to within a critical distance, backscattering occurs. Therefore, the tip in the upper half of the image visualizes the potential landscape of the lower QPC gate (which determines the trajectory of the lower edge). Equally, in the lower part of the SGM image, the potential landscape of the upper QPC gate is visualized.

From our data it is possible to estimate the widths $a_1$ and $a_2$ of the compressible strips at the edge of the sample and the width $a_{12}$ of the incompressible region separating these two channels~\cite{Aoki2005}. When the depletion spot induced by the tip brings two edge channels so close together that they start to interact, the conductance of the system decreases, until the two edges are completely backscattered. At this point two incompressible strips overlap, and this leads to a plateau in the conductance. Measuring the width of the region between the plateaus corresponding to $G = 4e^2/h$ and $G = 2e^2/h$,\footnote{As already mentioned, at $\nu_{bulk}=4$ two spin--degenerate edge channels are formed.} we obtain the following values: $322 \pm 57$~nm for the data in Fig.~\ref{fig:4}(a), $316 \pm 59$~nm for (b), $344 \pm 69$~nm for (c), and $360 \pm 38$~nm for (d). Within their error bars these values are equal, which gives an overall average value of $335 \pm 60$~nm for the width of two strips, i.e. the width of the inner edge channel is  $a_2 \approx 170$~nm. Interestingly, this value seems to be independent of the proximity of the outer edge. In Fig.~\ref{fig:4}(a), inner and outer edges propagate close together, while in Fig.~\ref{fig:4}(d) they are far from each other. Nevertheless, the width of the inner edge channel is the same within experimental uncertainty.  The data in Fig.~\ref{fig:4}(a) allow furthermore to measure the width $a_{12}$ of the incompressible strip which separates inner and outer edge channels ($2a_{12} = 75 \pm 12$~nm; $a_{12} \approx 40$~nm) and the width $a_1$ of the outer edge channels ($2a_1 = 176 \pm 29$~nm; $a_1 \approx 90$~nm). These last two values are somewhat larger than the values reported by Aoki et al.~\cite{Aoki2005}. On the other hand, our results are in good qualitative agreement with the theoretical description of Chklovskii et al.~\cite{Chklovskii1992}. According to these authors, the strip widths scale with the width of the depletion layer $l$ that separates the edge of the sample and the boundary of the 2DES as $l$ for compressible and as $l^{1/2}$ for incompressible strips. While Aoki et al. define their QPC by etching trenches in the sample, our samples have  metallic gates deposited on the sample surface. It seems reasonable to assume that etching induces a sharper potential step, which leads to a smaller $l$ and therefore to smaller values for the strip widths, in agreement with what was observed experimentally.

We have obtained very similar results for other bulk 2DES filling factors ($\nu_{bulk} = 2$, 2 spin--resolved edge channels; and $\nu_{bulk} = 6$, 3 spin--degenerate edge channels), which underlines the general validity of our findings to edge channels in the QH regime.

In conclusion, we have demonstrated a new method for the control of the trajectories of individual edge channels by the tip of the SGM.  Our results can represent a crucial  step for the implementation of multi--edge beam mixers and interferometers. We are now working on more refined device geometries to exploit this possibility for a control of the interaction of edge channels in QPC devices in the QH regime.


 \section*{Acknowledgements}
 \label{Acknowledgements}
 We thank R. M. Westervelt for comments and discussions. We acknowledge financial support from the Italian Ministry of Research (FIRB projects  RBIN045MNB  and RBID08B3FM).

\end{document}